\documentclass[12pt]{iopart}
\usepackage[dvips]{graphicx}
\usepackage{iopams,setstack,amssymb}

\newcommand{\E}{\mathrm{e}}
\newcommand{\I}{\mathrm{i}}

\newcounter{MaxMatrixCols}
\renewenvironment{pmatrix}{\left(\begin{array}{*{\value{MaxMatrixCols}}{c}}}{\end{array}\right)}%

\begin{document}
\title[Coulomb Gas Partition Function of a Layered Loop Model]
{Coulomb Gas Partition Function of a Layered Loop Model}

\author{Hirohiko Shimada}

\address{Department of Basic Sciences, University of Tokyo\\ 
Komaba 3-8-1, Meguro-ku, Tokyo 153-8902, Japan}
\ead{shimada@dice.c.u-tokyo.ac.jp}
\begin{abstract}
We consider a two-dimensional bi-layered loop model with a certain interlayer coupling and study its spectrum on a torus.
Each layer consists of an $O(n)$ model on a honeycomb lattice with periodic boundary conditions;
these layers are stacked such that the links of the lattice intersect each other.
A complex Boltzmann weight $\lambda$ with unit modulus
 is assigned to each intersection of two loops each from each layer. 
The model is reduced to an inhomogeneous vertex model at a special point of parameters. 
The continuum partition function is represented, based on the idea of the Coulomb gas, by a path integral over two compact bosonic fields.
The modular invariance of the partition function follows naturally.
Further, because of the topological nature of the interlayer coupling,  
the fluctuation of loops decomposes into a local and a global part. 
The existence of the latter leads to a sum over all the pairs of torus knots,
which can be Poisson ressummed by the M\"{o}bius inversion formula. 
This reveals the operator content of the theory.
The multiplicity of each operator is explicitly given by a combination of two Ramanujan sums.
We calculate each scaling dimension as a function of $\lambda$.  
We present the flow of dimensions which connects the decoupled-$O(1)$ models at $\lambda=1$ and the layered-$O(1)$ model with the non-trivial coupling  $\lambda=-1$. 
The lower spectrum in the latter model is related to that of a known coset model.
\end{abstract}
\section{Introduction}\label{section_introduction}

Statistical models in two dimensions 
can often be dealt with by several of the exact methods 
and have been invaluable sources of ideas in quantum field theory and many-body physics.  
However, little is known about two-dimensional models with layered structure,
which have been considered in various contexts \cite{ludwig, dotsenko, shimada, leclair, fendley}.
In this paper, we propose a bi-layered $O(n)$ loop model with a particular type of the interlayer coupling
which allows the exact calculation of the partition function by a simple path-integral based on the Gaussian action.

Our main tool is Coulomb gas (CG) method.
Let us recall some general features of CG.
As is well known, fluctuations of a system become important in the lower dimensions and, 
the critical exponents differ much from those of the Gaussian (or mean-field) theory.
Indeed, in two dimensions, we have a zoo of critical models which are quite different from the Gaussian;
for instance, they are the unitary minimal models \cite{fqs}, which satisfy highly-nontrivial bootstrap conditions 
under the infinite-dimensional conformal symmetry \cite{bpz}.
Interestingly,  however, it turned out that introducing intrinsic charges in a Gaussian theory 
can give heuristic, nevertheless exact, 
descriptions of a certain class of critical models 
\cite{nienhuis, dotsenkofateev,  cgon}.
Such descriptions are generically referred as the Coulomb gas
and enable us to calculate
the scaling dimensions \cite{nienhuis}, $N$-point functions \cite{dotsenkofateev}, and partition functions \cite{cgon}.

Let us turn to more specific lattice models whose continuum limits are described by the CG. 
The simplest is the $O(n)$ loop model 
(see \cite{jacobsen,nienhuisloop} for recent reviews).
This model actually have considerable importance
for general models with the central charge $c\leqslant 1$.
For instance, each member of the unitary minimal models can be represented as a superposition of several different loop models
\footnote{
More concretely, along the lines of \cite{pasquier, foda, cardy2}, one can see
the partition functions of the $O(n)$ model which is
generalized to have another parameter $\tilde{n}$ 
(a weight for non-contractible loops; see Section \ref{section_singlelayer})  
form a complete basis for those of the A-D-E lattice models ($\frac{1}{2}\leqslant c \leqslant 1$) \cite{pasquier}
and therefore for the minimal conformal field theories \cite{fqs, itzyksoncappelli,kato}. 
Further, the CG analysis of the loop models also provide us with the results on the non-unitary models such as 
the percolation, self-avoiding walk ($c=0$) \cite{cgon,saleur}, 
and dense polymer ($c=-2$) \cite{cgon,parisi}.}.

In a standard, algebraic view, 
partition functions of conformal field theories in general are assumed to have the 
sesquilinear form
\begin{eqnarray}
Z=\sum_{\{h,\bar{h}\}} N_{h,\bar{h}}\chi_{h}\chi_{\bar{h}},
\label{zchichi}
\end{eqnarray}
where $h$ is the allowed highest weight and
$\chi_{h}$ is the character of the representation in some algebra
\footnote{
The character formulas, for example, relevant to the unitary minimal models is that of the representation of the Virasoro algebra \cite{rocha}.};
an integer $N_{h,\bar{h}}$ is the multiplicity of the corresponding representation. 
This expression (\ref{zchichi}) reveals the operator content, from which one can extract various physics in the continuum limit.

The CG method allows us to evaluate 
the partition function of the $O(n)$ loop model on a torus in a remarkably direct way (without algebraic inputs).
It is essentially given by a path-integral 
over a compact bosonic field governed by the Gaussian action
\begin{eqnarray}
A[\varphi ]\propto \int d^2 z~ \left( \nabla \varphi(z)\right)^2,
\label{gaussianaction0}
\end{eqnarray}
where the bosonic field  $\varphi $ is a continuum version of the height function on the dual lattice 
which is uniquely determined from the non-intersecting loop configurations on a honeycomb lattice \cite{cgon}. 
The advantage of the CG methods is 
this kind of flexibility in scales compared to the other exact methods; 
from microscopic (lattice) formulation of a model, we can derive
the corresponding macroscopic (continuum) behavior,
which is yet straightforward to handle analytically.

In this paper, we extend the $O(n)$ loop model to a bi-layer model through the lattice formulation and 
calculate its continuum partition function on a torus in the form
by which one can, in general, read off the operator content.
The model has an interaction determined from 
the number of intersections between two loops each from each layer.
The decoupling of the local fluctuations from the topological ones,
which is essential in the single-layer loop model \cite{cgon},
is taken over to this model.
This is actually due to the topological nature of the inter-layer interaction.

As we mentioned, multi-layered models appear in many places.
In particular, it is important for the disordered systems \cite{ludwig, dotsenko, shimada}.
It is the essence of the replica method that
taking $M\to 0$ limit gives the physics of disordered system,
where $M$ is the number of layers.
Although it should be noted that our interlayer coupling might differ from that
arise from the replica method, we think this as a step towards 
the understanding of the replica limit.

As the principle of the path-integral dictates,
we should sum over all the topological sectors.
Accordingly, we obtain a sum over all the possible pairs of torus knots; 
each knot is characterized by two coprime integers and 
corresponds to a homotopical class of non-contractible loops on the one of the torus-layers.
From this expression, we can easily see the modular invariance \cite{cardy1} of 
the partition function.

To deal with the sum over the coprime integers,
we apply the M\"{o}bius inversion formula \cite{readsaleur}.
Our expression for the multiplicity in the bi-layer system
involves the product of the Ramanujan sums, and
looks similar to the one which occurs in 
the additive number theory
\footnote{More specifically, it resembles
the ``singular series" \cite{hardylittlewood} 
(see \cite{bogomolny,gadiyar} for physical application) which
is introduced to study both the (asymptotic) Waring's problem
and pair-correlation of two prime numbers. } .
It contains highly non-trivial selection rules.
To illustrate this, we analyze the partition function of the bi-layered O(1) models.

The rest of the paper is organized as follows.
In Section \ref{section_model},
the model is defined on the lattice 
with doubly-periodic boundary conditions.
At a special point of parameters, the model becomes
an inhomogeneous vertex model. 
We derive, in Section \ref{section_intersection}, a
determinant formula which counts the number of intersections.
In Section \ref{section_singlelayer}, we review the ideas used in calculating the partition function of the $O(n)$ model. 
In doing so, we fix the notations of the parameters used in the continuum formulation.
Section \ref{section_bilayer} contains our main result.
We first show the modular invariance of the partition function.
Next, we evaluate it by the repeated applications of 
the M\"{o}bius inversion formula.
The Poisson ressumation can then be performed, resulting in 
an expression for conformal dimensions of various operators, which
depends continually on the interaction parameter $\Gamma$.
Section \ref{section_bilayero1} is devoted to a brief 
inspection of the examples in the bi-layered $O(1)$ models. 
The lower spectrum seems to be related to  
that of a particular coset model. 
In Section \ref{section_flow}, we observe generic properties of the flow of the conformal dimensions
when one varies $\Gamma$.
In Section \ref{section_conclusion}, we conclude with few remarks.
In \ref{section_derivation}, a short derivation of the formula 
concerning
the M\"{o}bius inversion is given.
In \ref{section_isingorbifold}, the critical Ising parition function and its squares, which we need for checking one of our examples
are reviewed from a unified, orbifold point of view. 

\section{Lattice partition function of the layered loop model}\label{section_lattice}

\subsection{Interlayer coupling depending on the number of intersections}\label{section_model}
The $O(n)$ model is a spin model generalizing the Ising model, 
whose partition function, on a honeycomb lattice, allows the following loop representation:
\begin{eqnarray}
Z_{O(n);\; \mathrm{lattice} }
=\sum_{\mathrm{loops}} x^L n^C,
\label{zon}
\end{eqnarray}
where the sum is over non-self-intersecting loop configurations;
$L$ and $C$ is the length of loops and number of loops, respectively.
For $|n|\leq 2$, the model becomes critical at some  $x=x_c(n)$.
The universality class of the critical $O(n)$ model is studied well by
 mapping the model to 
 a six-vertex model on a Kagome lattice with the Coulomb gas method \cite{nienhuis}
or 
by mapping it to a seven-vertex model on the same honeycomb lattice, 
which is solved by the Bethe ansatz method \cite{baxter,batchelor}. 


A crucial step existing in both methods 
is to give an orientation to each loop.
Namely,  one decomposes the real weight $n$ in (\ref{zon}) into two complex weights:
\begin{eqnarray}
n=2\cos \pi\chi= n_++n_-,\ \mathrm{with}\ n_\pm=e^{\pm i\pi\chi},
\label{n2cos}
\end{eqnarray}
and assigns $n_+$ (resp. $n_-$) to each clockwise (resp. anti-clockwise) loop.
The partition function (\ref{zon}) is then written as
\begin{eqnarray}
Z_{O(n);\; \mathrm{lattice} } 
=\sum_{\substack{\mathrm{oriented}\cr \mathrm{loops}}} x^L n_+^{C_+}n_-^{C_-},
\label{zonoriented}
\end{eqnarray}
where the sum is over the configurations of oriented loops  and
$C_+$ (resp. $C_-$)  is the number of clockwise (resp. anti-clockwise) loops.
Using the relation 
$| \# \mathrm{left\ turns} -\#\mathrm{right\ turns}| =\pm 6$, 
which holds for the loops on the honeycomb lattice on a plane,
the weight $n_\pm$ can be further decomposed and redistributed as:
\begin{eqnarray}
e^{+i\pi\chi/6} \mathrm{(resp.}\ e^{-i\pi\chi/6}\mathrm{)
\ on\ each\ left\ (resp.\ right)\ turn,}
\label{onvertex}
\end{eqnarray}
 at a vertex, 
resulting in the seven-vertex model \cite{baxter,batchelor}.
\begin{figure}[!h]
\begin{center}
\includegraphics[width=6cm]{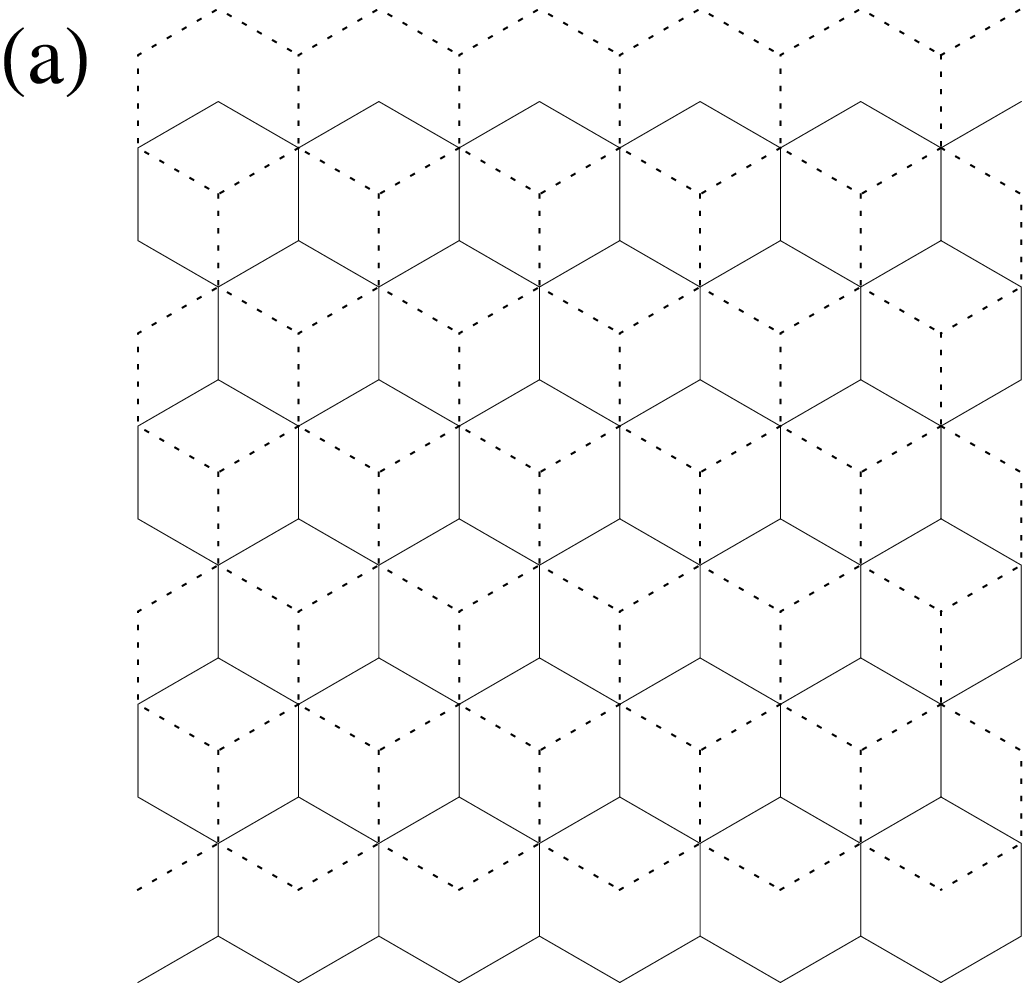}
\includegraphics[width=6cm]{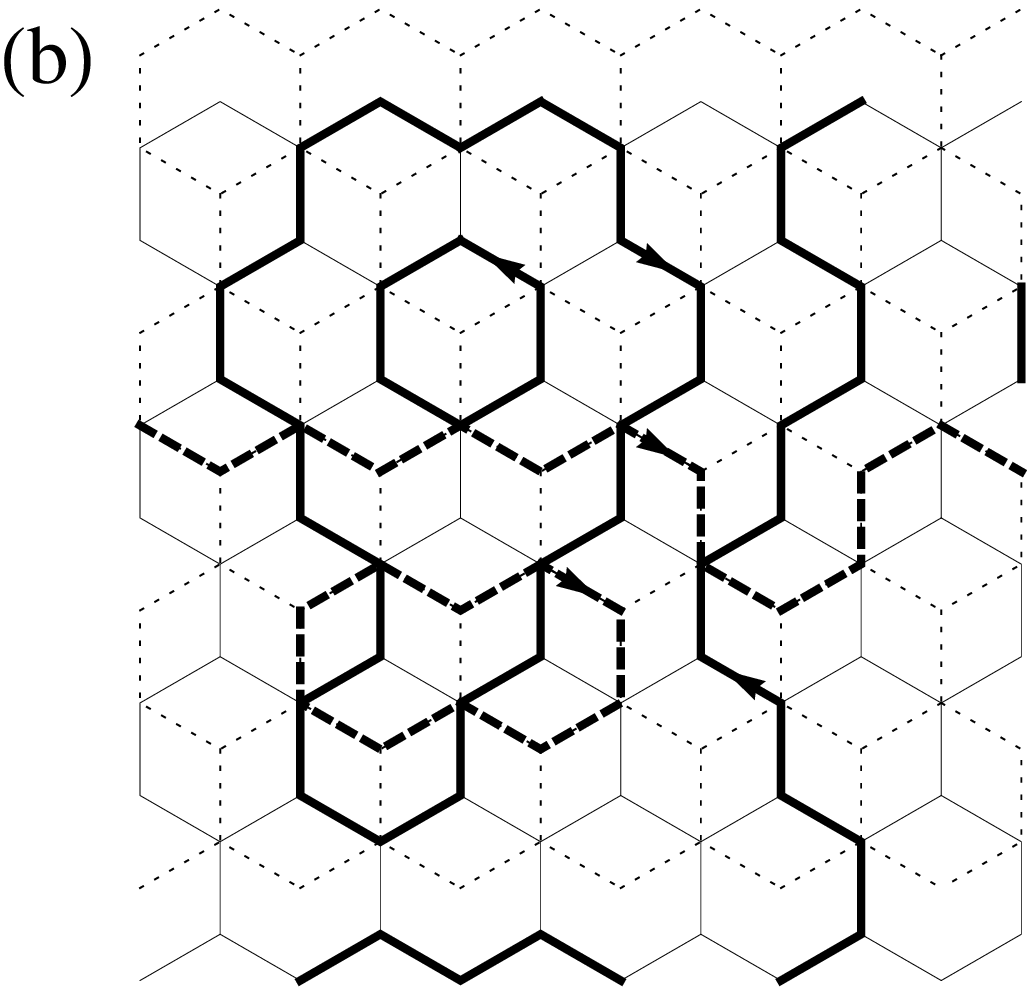}
\caption{
(a) The stacked lattice $A \bar{\sqcap} B$ formed by
the honeycomb lattice of the layer $A$ (solid lines) and 
that of the layer $B$ (dotted lines).
(b) A sample loop configuration on $A \bar{\sqcap} B$
endowed with the doubly-periodic boundary conditions.
This contains one non-contractible loop on each layer, and 
three contractible loops; two on $A$ and one on $B$. 
The total number of intersections $E$ equals $-1$.}
\label{loopconfig}
\end{center}
\end{figure}

We define the layered model via the summation (\ref{zonoriented}) over the oriented loops.
Consider two layers $A$ and $B$ each of which consists of 
the honeycomb lattice with doubly-periodic boundary conditions. 
To obtain an oriented loop configuration,
place arrows on some of the edges of $A$ and $B$ independently 
such that at each vertex there are 
either no arrows, or just one arrow in and just one out.
Now define a lattice formed by stacking 
$A$ and $B$ such that there are two types of the vertices, 
$V_3$ with three edges and $V_6$ with six edges;
we denote this by $A \bar{\sqcap} B$ (Fig. \ref{loopconfig}).
Then two oriented loops $l_A$ on $A$ 
and $l_B$ on $B$
can intersect each other on $A \bar{\sqcap} B$ at the vertex of type $V_6$. 
As an interlayer coupling, we assign a weight $\lambda^{\sigma}$, $(\sigma=\pm 1)$ 
for each intersection of loops depending on the orientations.
The sign $\sigma_P$ at an intersection P is defined locally and relatively for two oriented loops $l_A$ and $l_B$. 
Namely P is counted as $\sigma_P=+1$ (resp. $\sigma_P=-1$) if $l_B$ crosses $l_A$ at P from right to left (resp. from left to right). 

Let $E$ be the total number of signed intersections: $E=\sum_{i}\sigma_{P_i}$, where
$P_i$ is the $i$-th intersection. 
One easily verifies that $E=0$ if there are only loops 
each of which is homotopic to a point.  
Therefore, in order to obtain a coupled model, we have to consider 
the model on a manifold on which some of the loops can be non-contractible so that we may have $E\neq 0$; we here consider the model on a torus.

\begin{figure}[!h]
\begin{center}
\includegraphics[width=10cm]{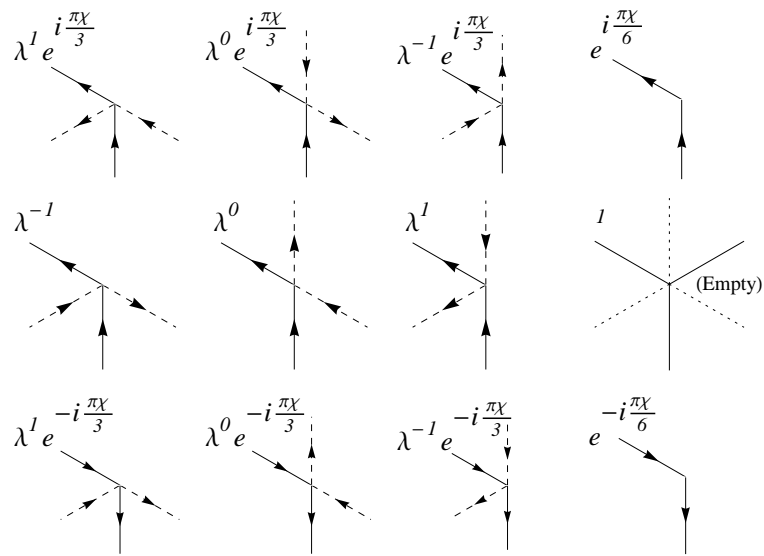}
\caption{Sample of the weights for arrow-configurations at each vertex of the type $V_6$. The solid lines and dotted lines 
(both with arrows) represent
the loop segments on the layer A and on the layer B, respectively. These weights are invariant under $2\pi/3$-rotations
of each configuration. 
Note that the weights at the top and bottom in the last column 
also appear in the set of weights at the vertex of the type $V_3$.}
\label{vertexweights}
\end{center}
\end{figure}

The partition function of the bi-layer model is then given by
\begin{eqnarray}
Z=\sum_{\substack{\mathrm{oriented}\cr \mathrm{loops}}} 
x^{L}n_+^{C_+}n_-^{C_-}\lambda^{E}
\label{zlat}
\end{eqnarray}
where $L$, $C_+$ (resp. $C_-$)
stand for a total length of loops, a number of contractible 
clockwise (resp. anti-clockwise) loops as before.
Since the interlayer coupling is defined locally,
the model (\ref{zlat}) can be thought of as a vertex model on the lattice $A \bar{\sqcap} B$.
The weight for an oriented-loop configuration on the lattice $A \bar{\sqcap} B$ is given by 
the product of the local vertex weights (Fig. \ref{vertexweights}).
This vertex model is special in its inhomogeneity:
the vertex of type $V_3$ allows $7$ arrow-configurations 
and
the vertex of type $V_6$ allows
$7^2=49$ configurations.
It would be nice if one could find the exact solution of this 
and compare that with an analysis based on our continuum partition function.

For an oriented non-contractible loop, we have
another relation $| \# \mathrm{left\ turns} -\#\mathrm{right\ turns}| =0$,
which is distinct from the relation for a contractible loop. 
Therefore, under the vertex rule (\ref{onvertex}), 
the non-contractible (oriented) loop in (\ref{zlat}) has the trivial weight $1$.
Instead, we can assign a different weight $\tilde{n}=\tilde{n}_++\tilde{n}_-$ with $\tilde{n}_+=\tilde{n}_-^*$: 
\begin{eqnarray}
Z=\sum_{\substack{\mathrm{oriented}\cr \mathrm{loops}}} 
x^{L}n_+^{C_+}n_-^{C_-}\tilde{n}_+^{D_+}\tilde{n}_-^{D_-}\lambda^{E},
\label{zlatnon}
\end{eqnarray}
where $D_+$ and $D_-$ are number of
positively and negatively oriented non-contractible loops, respectively.
For definiteness we use a convention in which 
we call ``positive'' for a loop of the winding numbers $(m_1,m_2)$
with $m_1>0$, or with $m_2>0$ if $m_1=0$.
In the continuum limit of the single layer model, 
this assignment of the weight is possible by 
a certain modification of the partition function \cite{cgon},
which will be used in Section \ref{section_continuum}.
In view of constructing the partition function basis (see the first footnote for single-layer),
it seems to be necessary to include this type of distinct weight $\tilde{n}$ in the definition.

Here we give a general consideration on 
the distinction of the physics according to the value of $\lambda$.
Consider an arbitrary loop configuration on the layer A, then
the weight for adding new non-contractible loop 
on the layer B will be, depending on the possible orientations,
$w_+=\tilde{n}_{+}\lambda^{E_0}$ or 
$w_-=\tilde{n}_{-}\lambda^{-E_0}$, 
where $E_0\in\mathbb{Z}$ is the additional intersection number.
If $|\lambda|=1$, we have $w_+=w_-^*$, and both orientation of the new loop occur 
in amplitudes with the same modulus.
On the other hand, if $|\lambda|\neq 1$, 
this symmetry is violated; 
one of the two possible orientation is preferred to the other.
In this paper, we only consider  
the ``symmetric'' case: $|\lambda|=1$.
Thus, we set
\begin{eqnarray}
\lambda=\E^{2\pi i \hat{g} \Gamma},
\label{lambda}
\end{eqnarray}
for $\Gamma\in\mathbb{R}$. 
The scale of the parameter $\Gamma$
is chosen for the later convenience,
and the rescaled coupling $\hat{g}$ will be defined in (\ref{ghat}).

\subsection{Counting the number of intersections between non-contractible loops}\label{section_intersection}

\begin{figure}[!h]
\begin{center}
\fl
\includegraphics[width=11cm]{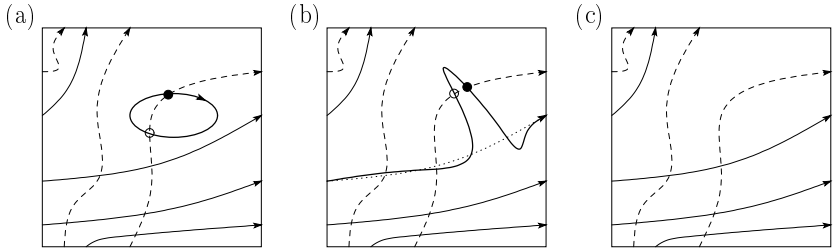}
\caption{
(a) An addition of one contractible loop.
(b) A local deformation of a loop.
(c) Before the deformations.
The opposite sides of the rectangle are identified.
Since intersections indicated by the black and white dots have relatively opposite signs,
the total number of intersections $E$ is common (equal to five) for all the three cases. }
\label{localdeformations}
\end{center}
\end{figure}

Our purpose in introducing the signs to the intersections is 
to set the weight induced by the interlayer-coupling    
independent on a local fluctuation of loops.
This is guaranteed by the
invariance of the intersection number $E$ under   
\begin{itemize}
\item (i) an addition of contractible loops (Fig. \ref{localdeformations}-(a)), 
\item (ii) a local deformation of loops (Fig. \ref{localdeformations}-(b)).  
\end{itemize}
Actually, it is these properties that enable us to decompose the bosonic field into local and global part
and to do a path integral calculations in Section \ref{section_continuum}. 

Now let us count the total intersection numbers $E$ between loops belonging to the distinct layers.
Firstly, one may forget about contractible loops by the invariance under (i). 
The result can be stated in terms of the pair of total winding numbers
$(l_A,-m_A)$  and $(l_B,-m_B)$ \footnote{The reason for the minus before $m$ will become clear when we define the frustration in Section \ref{section_singlelayer}.}
due to the presence of the several non-contractible loops on each layer.
Below, we show the following:
\begin{eqnarray}
E&(l_A,-m_A;\; l_B, -m_B)=\det\begin{pmatrix}m_A&m_B \\ l_A&l_B\end{pmatrix}.
\label{determinant}
\end{eqnarray}
Since loops on a honeycomb lattice should be non-self-intersecting, 
only one type of homotopy class is allowed on each layer 
(otherwise they intersect within the same layer).
This reduces the problem to 
the calculation of the intersection number between more fundamental objects.
Namely, they are the torus knots characterized respectively by 
the winding numbers $(l'_A,-m'_A)$  and $(l'_B,-m'_B)$ which are coprime
\begin{equation} 
m'_A\wedge l'_A=m'_B\wedge l'_B=1, 
\label{coprime}
\end{equation}
where $m \wedge l$ stands for the greatest common divisor 
(GCD) of $m$ and $l$;
one should merely multiply $ab$ to $E(l'_A,-m'_A;\; l'_B, -m'_B)$,
if the configuration consists of the total winding number with 
$m_A\wedge l_A=a$ and $m_B\wedge l_B=b$.
This procedure is compatible with the bi-linearity of the determinant.
Thus, we can assume (\ref{coprime}) without loss of generality.

To do the rest of the calculation, we consider an infinite plane with a coordinate $(\xi, \eta)$ in which the points differing by integer coordinates are identified and use it to parametrize the knots.
Each of them is then represented by a ``closed" curve in the unit square $(\xi, \eta)\in U\equiv [0,1]\times [0,1]$.
By the invariance under (ii), we may stretch the curve 
to form a set of pararell lines; these are periodic geodesics on 
the torus.
We locate these to pass through the origin.
Then the locations of the intersections in $U$ and their mirrors outside $U$
are given by the solutions 
$(\xi, \eta)$ of
\begin{eqnarray}
m'_A(\xi-\xi_0)+l'_A(\eta-\eta_0)&=0\\
m'_B(\xi-\tilde{\xi}_0)+l'_B(\eta-\tilde{\eta}_0)&=0,
\end{eqnarray}
where each set of $(\xi_0, \eta_0)\in \mathbb{Z}^2$ and $(\tilde{\xi}_0, \tilde{\eta}_0)\in \mathbb{Z}^2$ defines two lines and their unique intersection point.
Since, by the B\'{e}zout's lemma, 
(\ref{coprime})
implies that
there exists a pair of integers $(\xi_0, \eta_0)\in \mathbb{Z}^2$ 
such that $m'_A \xi_0+l'_A \eta_0=\alpha$ for any integer $\alpha$
and so does $(\tilde{\xi}_0, \tilde{\eta}_0)\in \mathbb{Z}^2$ such that $m'_B \tilde{\xi}_0+l'_B \tilde{\eta}_0=\beta$ for any integer $\beta$, one can equivalently count the solutions of,
\begin{eqnarray}
\begin{pmatrix}
m'_A & l'_A \cr m'_B & l'_B
\end{pmatrix}
\begin{pmatrix}
\xi   \\ \eta
\end{pmatrix}
=
\begin{pmatrix}
\alpha \\ \beta
\end{pmatrix}
\in \mathbb{Z}^2.  
\label{matrixeq}
\end{eqnarray}
The set of the solutions $(\xi,\eta)$ forms a lattice,
which can be linearly mapped from 
the unit square lattice $\mathbb{Z}^2$ by the multiplication of the inverse of the matrix in (\ref{matrixeq}) . 
By taking the Jacobian factor into account, 
one can count the number of the solutions in $U$.
We have thus shown the formula (\ref{determinant}).

\begin{figure}[!h]
\begin{center}
\includegraphics[width=10cm]{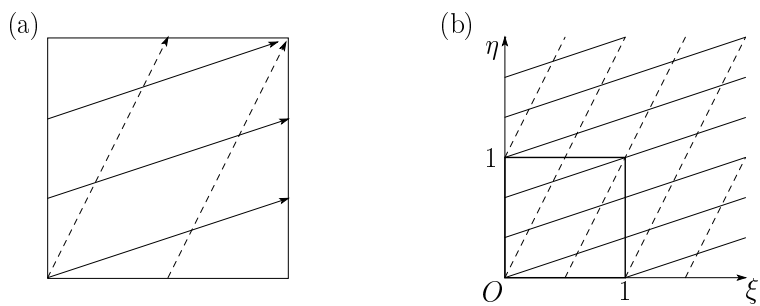}
\caption{
(a) The streched configuration that is topologically connected with the configuration in Fig. 
\ref{localdeformations}-(c).
Again the number of intersections is five; 
the intersection at the top-right and bottom-left
shoud be counted as one-half for each because of the identification. 
The winding numbers are given by $(-l_A,m_A)=(3,1)$
and $(-l_B,m_B)=(1,2)$.
(b) The fundumental region $U$ and its mirror on $(\xi,\eta)$-plane.
}
\label{xieta}
\end{center}
\end{figure}

\section{Continuum partition function}\label{section_continuum}

We first recall the Coulomb gas partition function 
of the $O(n)$ model \cite{cgon}.
Using the determinant formula (\ref{determinant}),
we formulate the continuum partition function of the layered model (\ref{zlatnon}).
Here the moduli parameter of the torus is denoted by
$\tau=\omega_2/\omega_1=x+i y$ where 
$\omega_1$ and $\omega_2$ are the two basic periods. 

\subsection{Single layer} \label{section_singlelayer}

We introduce a compactified bosonic field $\varphi(z)$ governed by the action 
\begin{eqnarray}
A[\varphi ]&=\frac{g}{4\pi}\int d^2 z~ \left( \nabla \varphi(z)\right)^2.
\label{gaussianaction}
\end{eqnarray}
This field can be considered as a continuum version of the height function $h(r_i)$; 
the latter is defined on the dual lattice $\{r_i\}$ and changes its value by $\phi_0=\pm 2\pi f$ when $r_i$ goes across the loop.
The sign is again dependent on the orientation of the crossing; 
it is determined in a way parallel to the definition of $\sigma_P$ below (\ref{onvertex}).
Since loops are closed, the correspondence between the loop and height configuration is unique
up to a constant shift of the latter.
On the torus, a configuration of the field $\varphi$ may belong to
non-trivial topological sector called
the ``frustrations'' along the periods $(\omega_1$, $\omega_2)$,
which are denoted here by 
$(\delta_1 \varphi,\delta_2 \varphi)=2\pi f (m,l)$ with $(m,l)\in\mathbb{Z}^2$.
Now, in the continuum limit, the partition function (\ref{zonoriented}) on a torus
becomes 
\begin{eqnarray}
Z_{\mathrm{gauss}}(g,f)&=\sum_{m,l}Z_{m,l}(g,f), \label{gaussian}\\
Z_{m,l}(g,f)&=\int_{(\delta_1 \varphi,\delta_2 \varphi)=2\pi f(m,l)}\hspace{-15 mm}
\mathcal{D}\varphi~ e^{-A[\varphi]}=Z_0\left(gf^2\right)\exp\left(-\frac{\pi g f^2 |m\tau-l|^2}{y}\right),
\label{zml}
\end{eqnarray}
where the field $\varphi$ is divided into global and local part: $\varphi=\varphi_{m,l}+\varphi_0$.
Quite importantly, the local fluctuation $\varphi_0$ can be integrated out, giving $Z_0(g)$ as \cite{itzykson}
\begin{eqnarray}
Z_0(g)&=\int\mathcal{D}\varphi_0~ e^{-A[\varphi]}
=\sqrt{\frac{g}{y}}|\eta(q)|^{-2},
\label{z0}
\end{eqnarray}
where we have used the variable $q=\E^{2\pi i \tau}$ and the Dedekind eta function
\begin{eqnarray}\label{eta}
\eta(q)=q^{\frac{1}{24}}\prod_{n=1}^\infty \left(1-q^n\right).
\end{eqnarray}
In the standard convention, $\phi_0=\pi$ $(f=1/2)$;
the coupling $g$ is determined by 
\footnote{The value of $g$ is fixed from the requirement that the screening field 
$\E^{i \varphi/f}$,
which originates from the discreteness of the height function
$h(r_i)\in 2\pi f \mathbb{Z}$, be marginal \cite{kondev}.
This leads to the relation $g=1\pm\chi$ with $\chi$ defined in (\ref{n2cos}).
}   
\begin{eqnarray}\label{ng}
n=-2\cos \pi g .
\end{eqnarray}
Such description by the local action (\ref{gaussianaction})
corresponds to the vertex rule (\ref{onvertex}).
Thus, the weight for the non-contractible loop (without orientation) is automatically
$\tilde{n}=2$.
A novel observation \cite{cgon} is that
one can modify the weight for non-contractible loop as
\begin{eqnarray}
\tilde{n}=2\longrightarrow \tilde{n}=2\cos \pi e_0
\end{eqnarray} 
by exploiting the identity: 
\begin{eqnarray}
(2 \cos \pi e_0)^M
=\prod_{j=1}^M\left( \sum_{\varepsilon_j=\pm1}\E^{i\pi e_0  \varepsilon_j}\right)
=\sum_{\{\varepsilon_j\}} \cos{\left(\pi e_0 
\sum_{j=1}^M \varepsilon_j\right)},
\end{eqnarray}
where $\{\varepsilon_j\}$ represents $2^{M}$ ways of the positive or the negative orientation
of $M$ non-oriented non-contractible loops.
A presence of the loop which is characterized by the coprime winding numbers 
$(l',-m')$
leads to the frustration 
$(\delta_1 \varphi,\delta_2 \varphi)=2\pi f (m',l')$.
Thus, the net number of oriented loops is related to the frustration as
\begin{eqnarray}
\left|\sum_{j=1}^M \varepsilon_j\right|=m\wedge l.
\end{eqnarray}
As the result, the partition function is given by \cite{cgon}
\begin{eqnarray}
Z_{O(n)}&=\sum_{m,l}Z_{m,l}\left(\hat{g}\right) \cos \left(\pi e_0 m\wedge l\right),
\label{zongcd}
\end{eqnarray}    
where we have introduced, for convenience, the rescaled coupling 
\begin{eqnarray}\label{ghat}
\hat{g}=\frac{g}{4},
\end{eqnarray}    
and have used short-hand notation 
$Z_{m,l}\left(\hat{g}\right)=Z_{m,l}\left(\hat{g},1\right)$
and the relation $Z_{m,l}\left(gf^2,1\right)=Z_{m,l}\left(g,f\right)$
with $f=1/2$.

\subsection{Bi-layer} \label{section_bilayer}

From the lattice expression (\ref{zlatnon}), definition
(\ref{lambda}), and formula (\ref{determinant}),
the partition function of the layered $O(n)$ model can be written as
\begin{eqnarray}
Z=\sum_{m_A,l_A}\sum_{m_B,l_B}&Z_{m_A,l_A}\left(\hat{g}\right)
\cos \left(\pi e_0 m_A\wedge l_A\right)C(m_A,l_A,m_B,l_B;\Gamma)\nonumber\\
&\cdot Z_{m_B,l_B}\left(\hat{g}\right)\cos \left(\pi e_0 m_B\wedge l_B\right) 
\label{zlayer}
\end{eqnarray}
where the inter-layer coupling is given by
\begin{eqnarray}
C&(m_A,l_A,m_B,l_B;\Gamma)=\exp \left[2\pi i \hat{g}\Gamma \cdot\det\begin{pmatrix}m_A&m_B \\ l_A&l_B\end{pmatrix}\right].
\label{cdeterminant}
\end{eqnarray}
At the level of the action, this term would correspond to the coupling 
\footnote{The author thanks J. Cardy and J.L. Jacobsen for comments concerning this point.
The author also notes that a similar (but different) action was used to study the
conformal theories which posses $\mathcal{W}_n$ symmetry \cite{kostov}.}
\begin{eqnarray}
S_{AB}[\varphi ]&=-i\frac{ g\Gamma}{2\pi}\int d^2 z~ 
\vec{h}\cdot\left( \nabla \varphi_A(z) \times \nabla \varphi_B(z)\right),
\label{couplingaction}
\end{eqnarray}
where $\vec{h}$ is the positive unit vector perpendicular to the $z$-complex plane.
Since the integrand has the canonical dimension two, a possibility is that
the presence of the inter-layer coupling leads to a marginal deformation of 
the decoupled model ($\Gamma=0$).

In general, an invariance of partition functions under the modular transformation:
\begin{eqnarray}
\tau\to\frac{a\tau +b}{c\tau +d},\quad\quad \begin{pmatrix}
a&b\\
c&d
\end{pmatrix}\in SL(2, \mathbb{Z}),
\label{modular}
\end{eqnarray}
imposes strong constraints on the operator content \cite{cardy1}.  
In this respect, 
our expression (\ref{zlayer}) of the 
bi-layer system automatically satisfies the modular invariance.
To see this, first recall that any modular transformation (\ref{modular}) can be
expressed as a
finite composition of 
the two basic transformations $T: \tau \to \tau+1$ and $S: \tau \to -1/\tau$.
The modular invariance of (\ref{gaussian}) 
follows from the transformation properties of the frustrations 
($\hat{g}$-dependence being suppressed):
\begin{eqnarray}
T:\:Z_{m,l}\left(\tau+1\right)&=Z_{m,l-m}\left(\tau\right),\\
S:\:Z_{m,l}\left(-\frac{1}{\tau}\right)&=Z_{-l,m}\left(\tau\right),
\end{eqnarray}
and from the basic idea of the path-integral that 
the sum should be taken over all the possible configurations 
(here, taken over all the frustrations).
Since the GCD remains invariant under these transformation: 
\begin{equation}
m\wedge l=m\wedge (l-m)=(-l)\wedge m,
\end{equation}
the partition function of the $O(n)$ model (\ref{zongcd}) also has
the modular invariance.
Further, the determinant has the analogous invariance,\footnote{The invarince of the determinant under the general modular transformation
can be also seen by the fact that the frustration vectors
$(m_A,l_A)$ and $(m_B,l_B)$ transform
by multiplication of the identical $SL(2,\mathbb{Z})$ matrix, 
which has the determinant equal to one.}
which may be 
represented as 
\begin{equation}
\vec{M}\times \vec{L}=\vec{M}\times (\vec{L}-\vec{M})=(-\vec{L})\times \vec{M}
\end{equation}
if the determinant in (\ref{cdeterminant}) is regarded as
the third component of $\vec{M}\times \vec{L}$
with $\vec{M}=(m_A,m_B ,0)$ and $\vec{L}=(l_A,l_B ,0)$.
Thus, under the transformations,
each term in the sum (\ref{zlayer})
changes into 
another term with the common extra factor, 
compared with (\ref{gaussian}),
determined only from the GCD and determinant of the frustrations.
This establishes the modular invariance of the partition function
of the layered model (\ref{zlayer}).

In order to see the operator content,
we need to Fourier transform the sum (\ref{zlayer})
into a series in $q$ and $\bar{q}$.
The similar transformation for the partition function (\ref{gaussian})
is easily done by the Poisson ressumation.
Intuitively, this task becomes rather non-trivial 
for the sum with the GCD, (\ref{zongcd}) and (\ref{zlayer}),
because the GCD function $m\wedge l$ 
has no apparent periodic structure on the $(m,l)$-plane;
its distribution has rather recursive (self-similar) and irregular pattern
(being recursive because all the prime numbers play equivalent and relatively independent roles
in determining the GCD,
and irregular because they appear irregularly). 
It is possible, however, to take advantage of the recursive nature by allowing deliberate overlaps and to express the sum (\ref{zongcd})
as a superposition of sums over
the periodic structures $(m,l)\in k\mathbb{Z}^2$.
In order to do this, we use the following formula:
\begin{eqnarray}\label{mobius}
\fl
\sum_{m,l}Z_{m,l}\left(\hat{g}\right)
\cos \left(\pi e_0 m\wedge l\right)
&=&\sum_{d>0}\sum_{k:~d|k}\mu\left(\frac{k}{d}\right)\cos \left(\pi e_0 d\right)
\sum_{k|(m,l)}Z_{m,l}
\end{eqnarray}
where the M\"{o}bius function is defined by
\begin{eqnarray}
\fl
\mu (n)=
\cases{
1 & {\rm if $n=1$,}\\
(-1)^k & {\rm if $n$ is a square-free integer with $k$ distinct prime factors,} \\
0 & {\rm if $n$ is not square-free.}
}
\end{eqnarray}
Employing such technique based on
a version of the M\"{o}bius inversion formula \cite{hardy},
the partition function of the $O(n)$ model was evaluated \cite{readsaleur}.
Changing the order of summation over $d$ and $k$ in (\ref{mobius}), and then
using that formula twice, we obtain
\begin{eqnarray}
\fl
Z=\sum_{k_A>0,\,k_B>0}\sum_{d_A|k_A,\,d_B|k_B}
&\mu\left(\frac{k_A}{d_A}\right)
\mu\left(\frac{k_B}{d_B}\right)
\cos \left(\pi e_0 d_A\right)
\cos \left(\pi e_0 d_B\right)\nonumber\\ 
&\cdot
\sum_{k_A|(m_A,l_A)}
\sum_{k_B|(m_B,l_B)}Z_{m_A,l_A}C(m_A,l_A,m_B,l_B;\Gamma)
Z_{m_B,l_B}.
\label{mumucoscoszcz}
\end{eqnarray}
Note that the sum in the second line factorizes as below:
\begin{eqnarray}
\fl
Z_0(\hat{g})^{-2}&\hspace{-11mm}\sum_{k_A|(m_A,l_A)}\sum_{k_B|(m_B,l_B)}
Z_{m_A,l_A}C(m_A,l_A,m_B,l_B;\Gamma)Z_{m_B,l_B}\nonumber\\
&=
\sum_{\substack{l_A:~k_A|l_A,\cr l_B:~k_B|l_B}}
\E^{-\frac{\pi \hat{g}}{y} \left[(m_Ax+l_A)^2+m_A^2y^2\right]}
\E^{2\pi i\hat{g} \Gamma (m_A l_B-l_A m_B)}
\E^{-\frac{\pi \hat{g}}{y} \left[(m_Bx+l_B)^2+m_B^2y^2\right]}\nonumber\\
&= S_{k_A}^{(A)}(-m_B)S_{k_B}^{(B)}(+m_A),
\label{zczss}
\end{eqnarray}
where the last two factors are given by
\begin{eqnarray}
S_k^{(A)}(-m_B)=&\sum_{l_A:~k|l_A} \E^{-\frac{\pi \hat{g}}{y} \left[(l_A+m_Ax)^2+m_A^2y^2\right]+2\pi i\hat{g}\Gamma l_A\cdot(-m_B)},\\
S_k^{(B)}(+m_A)=&\sum_{l_B:~k|l_B} \E^{-\frac{\pi \hat{g}}{y} \left[(l_B+m_
Bx)^2+m_B^2y^2\right]+2\pi i\hat{g}\Gamma l_B\cdot(+m_A)}.
\end{eqnarray}
Then we shall perform the Poisson resummation to each quantity.
It is sufficient to see this for, say $S_k^{(A)}(-m_B)$; this goes as: 
\begin{eqnarray}
\fl
S_k^{(A)}(-m_B)
&=&\sum_{l_A:~k|l_A} \E^{-\frac{\pi \hat{g}}{y} 
\left[(l_A+m_Ax + i\Gamma m_By)^2 - 2i\Gamma m_Am_Bxy+(\Gamma m_By)^2+m_A^2y^2\right]}\nonumber\\
&=&\sqrt{\frac{y}{\hat{g}k^2}}\sum_{p\in \mathbb{Z}} \E^{-\pi y 
\left[\frac{p^2}{\hat{g}k^2} +\hat{g} (\Gamma m_B)^2+\hat{g} m_A^2\right]}
\E^{2 \pi i 
\left(\frac{m_Ax + i\Gamma m_By}{k}\right)p}
\E^{2 \pi i \hat{g}\Gamma m_Am_Bx}\nonumber\\
&=&\sqrt{\frac{y}{\hat{g}k^2}}\sum_{p\in \mathbb{Z}} \E^{-\pi y 
\left[\frac{p^2}{\hat{g}k^2} +\hat{g} (\Gamma m_B)^2+\hat{g} m_A^2 + 2\frac{\Gamma m_Bp}{k}\right]}
\E^{2 \pi i x
\left(\frac{m_Ap}{k}+\hat{g}\Gamma m_Am_B\right)}
\label{poisson_nsum}
\end{eqnarray}
Comparing this with the expression
$q^h\bar{q}^{\bar{h}}=\exp\left[-2\pi y(h+\bar{h})+2\pi i x(h-\bar{h})\right]$, we obtain
\begin{eqnarray}
S_k^{(A)}(-m_B)=\sqrt{\frac{y}{\hat{g}k^2}}\sum_{p\in \mathbb{Z}}
q^{h(k,p,m_A,m_B)}\bar{q}^{\bar{h}(k,p,m_A,m_B)},
\label{sqq}
\end{eqnarray}
with 
\begin{eqnarray}
h(k,p,m_A,m_B)=&\frac{1}{4}
\left(\frac{p}{\sqrt{\hat{g}}k}+\sqrt{\hat{g}}\left(m_A + \Gamma m_B\right)\right)^2,\\
\bar{h}(k,p,m_A,m_B)=&\frac{1}{4}
\left(\frac{p}{\sqrt{\hat{g}}k}-\sqrt{\hat{g}}\left(m_A - \Gamma m_B\right)\right)^2.
\end{eqnarray}
The magnetic charge of the layer $A$ and that of $B$ are mixed
by the presence of the interlayer coupling $\Gamma$.
If we define
\begin{eqnarray}
\fl
\Delta(p_A/k_A,p_B/k_B,m_A,m_B)&=&
h(p_A/k_A,m_A,m_B)+h(p_B/k_B,m_B, -m_A),
\label{delta}\\
\fl
\bar{\Delta}(p_A/k_A,p_B/k_B,m_A,m_B)&=&
\bar{h}(p_A/k_A,m_A,m_B) + \bar{h}(p_B/k_B,m_B, -m_A),
\label{deltabar}
\end{eqnarray}
then the scaling dimension and conformal spin yield
\begin{eqnarray}
\fl
\Delta+\bar{\Delta}=&
\frac{1}{2}\left[\frac{1}{\hat{g}}\left(\frac{p_A^2}{k_A^2}+\frac{p_B^2}{k_B^2}\right) 
+\hat{g}\left( 1+\Gamma^2\right)\left( m_A^2+m_B^2\right)\right]+
\left(\frac{p_A}{k_A}m_B - m_A\frac{p_B}{k_B}\right)\Gamma
\label{cdimension}\\
\fl
\Delta-\bar{\Delta}=&
\frac{p_Am_A}{k_A}+\frac{p_Bm_B}{k_B},
\label{cspin}
\end{eqnarray}
where the arguments of 
$\Delta$ and $\bar{\Delta}$ are the same as the one written in 
(\ref{delta}) and (\ref{deltabar}).

Substituting equations (\ref{z0}), (\ref{zczss}), and (\ref{sqq})
into (\ref{mumucoscoszcz})
we obtain
\begin{eqnarray}
\fl
|\eta(q)|^4 Z=\sum_{k_A>0,\,k_B>0}&\sum_{d_A|k_A,\,d_B|k_B}
\frac{\mu\left(\frac{k_A}{d_A}\right)
\mu\left(\frac{k_B}{d_B}\right)}{k_Ak_B}
\cos \left(\pi e_0 d_A\right)
\cos \left(\pi e_0 d_B\right)\nonumber\\ 
&
\sum_{\substack{m_A:~k_A|m_A,\cr m_B:~k_B|m_B}}
\sum_{\substack{p_A\in \mathbb{Z}, \cr p_B\in \mathbb{Z}}}
q^{\Delta}\bar{q}^{\bar{\Delta}},
\label{zqq}
\end{eqnarray}
In order to see the operator content,
we should evaluate the coefficient of
$q^{\Delta}\bar{q}^{\bar{\Delta}}$ with fixed sets of
values $(\Delta, \bar{\Delta})$. 
Since these values depend on the electric charges $e_i=p_i/k_i$ $(i=A,B)$ which are invariant under
$(p_i, k_i) \to (c p_i, c k_i)$ for positive integer $c$, 
we would like to group these terms, which gives the same values of $(\Delta, \bar{\Delta})$.

For this purpose,
it is natural to decompose the sum (\ref{zqq})
into terms with $m_Am_B\neq 0$ and $m_Am_B= 0$,
which are respectively denoted by $\mathcal{Z}_{\neq 0}$ and $\mathcal{Z}_0$ so that
\begin{eqnarray}
|\eta(q)|^4 Z=\mathcal{Z}_{\neq 0}+\mathcal{Z}_0,
\label{zzz}
\end{eqnarray}
For the sum $\mathcal{Z}_{\neq 0}$, we apply
the formula derived in \ref{section_derivation}:  
\begin{eqnarray}
\fl
\sum_{k>0}\sum_{d|k}
\frac{\mu\left(\frac{k}{d}\right)}{k}
\cos \left(\pi e_0 d\right)&
\sum_{m:~k|m, m\neq0}
\sum_{p\in \mathbb{Z}}
f\left(\frac{p}{k},m\right)\nonumber\\
\fl&=\sum_{m\neq 0}\sum_{k:~k|\hat{m}}
\Lambda\left(k,\hat{m}\right)
\sum_{p:~p\wedge k=1}
f\left(\frac{p}{k},m\right)
\label{appramanujan}
\end{eqnarray}
where $f\left(\frac{p}{k},m\right)$ is an arbitrary function (thus, not necessarily even in the variable $m$); the notation $\hat{m}=|m|$ is used for readability. 
Here, $\Lambda\left(k,m\right)$ is the multiplicity in the 
single layer $O(n)$ model given by the finite sum: 
\begin{eqnarray}
\Lambda\left(k,m\right)=\hat{m}^{-1}
\sum_{d:~d|\hat{m}}
\mathcal{C}_{\frac{\hat{m}}{d}}\left(\frac{\hat{m}}{k}\right)\cos \left(\pi e_0 d\right),
\label{lamda}
\end{eqnarray}
where 
$\mathcal{C}_a(b)$ is the Ramanujan sum defined by
\begin{eqnarray}
\mathcal{C}_a(b)\equiv\sum_{\substack{1\leqslant h\leqslant n\cr h\wedge a=1}} \E^{2\pi\I \frac{b}{a}h}.
\end{eqnarray}
Then we have
\begin{eqnarray}
\mathcal{Z}_{\neq 0}=
\prod_{i=1,2}
\left[\sum_{m_i\neq 0}\sum_{k_i:~k_i|\hat{m}_i}\Lambda\left(k_i,m_i\right)\sum_{p_i:~p_i\wedge k_i=1}\right]
q^{\Delta}\bar{q}^{\bar{\Delta}
\left(\frac{p_1}{k_1},\frac{p_2}{k_2},m_1,m_2\right)}.
\label{zoff}
\end{eqnarray}
Next, we consider the sum $\mathcal{Z}_{0}$
which consists of the terms 
with which at least one of $m_A$ and $m_B$ vanishes.
Taking the terms with $m_B=0$ and $m_A\neq 0$, for example, 
we see from (\ref{zlayer}) that the sum over $l_A$ reduces to  
the nonzero-magnetic-charge 
sector of the single layer $O(n)$ model,
which can be evaluated using 
the formula (\ref{mobius}).
Further, since 
the GCD is to be evaluated as
$0\wedge l_B=l_B$, 
the sum over $l_B$ is just
\begin{eqnarray}
\fl
\frac{1}{2}\sum_{s=\pm1}\sum_{l_B} \E^{-\frac{\pi \hat{g}}{y} l_B^2
+2\pi i\hat{g}\left(s \frac{e_0}{2\hat{g}}+\Gamma m_A\right)l_B}
=\sqrt{\frac{y}{4\hat{g}}}\sum_{s=\pm1}\sum_{p_B}
\E^{-\frac{\pi y}{\hat{g}} \left(p_B+s\frac{e_0}{2} +\hat{g} \Gamma m_A \right)^2}.
\end{eqnarray}
Therefore we see that setting $m_B=0$ amounts to
a restriction on the electoric charge 
in (\ref{cdimension}) and (\ref{cspin})
to $e_B=p_B/k_B$ with 
$k_B=1$ and a shift in $p_B\to p_B\pm\frac{e_0}{2} +\hat{g} \Gamma m_A$. 
Taking the terms with $m_A=0$ and 
the symmetry
$\Lambda\left(k,m\right)=\Lambda\left(k,-m\right)$
into account, we obtain
\begin{eqnarray}
\mathcal{Z}_{0}=\sum_{p_1,p_2}&q^{\Delta}\bar{q}^{\bar{\Delta}
\left(p_1+\frac{e_0}{2},p_2+\frac{e_0}{2},0,0\right)}\nonumber\\
&+2\sum_{m\neq 0}\sum_{k:~k|\,|m|}
\Lambda\left(k, m\right)
\sum_{p_1:~p_1\wedge k=1}
\sum_{p_2}q^{\Delta}\bar{q}^{\bar{\Delta}
\left(\frac{p_1}{k},p_2+\frac{e_0}{2}+\hat{g}\Gamma m,m,0\right)},
\label{zdiagoff}
\end{eqnarray}
To summarize, our result on the partition function of the bi-layer model 
is given by (\ref{zzz}) with  (\ref{zoff}) and (\ref{zdiagoff}).
 
\setcounter{footnote}{1}

\section{Spectrum of the bi-layer models}

The first rather trivial observation is the following.
When the inter-layer coupling $\Gamma$ vanishes,
the scaling dimension (\ref{cdimension}) becomes just a sum of 
two dimensions known in usual CG.
Hence,
the bi-layer model (\ref{zqq}) decouples into two independent $O(n)$ models:
\begin{eqnarray}
Z(\hat{g}, e_0, \Gamma=0)=&\left[Z_{O(n)}(\hat{g}, e_0)\right]^2
\end{eqnarray}
with $Z_{O(n)}(\hat{g}, e_0)$ given by \cite{jacobsen}
\begin{eqnarray}
|\eta(q)|^2Z_{O(n)}(\hat{g},e_0)=&\sum_{p_1}q^{\Delta}\bar{q}^{\bar{\Delta}
\left(p_1+\frac{e_0}{2},0,0,0\right)}\nonumber\\
&+\sum_{m\neq 0}\sum_{k:~k|\,|m|}
\Lambda\left(k, m\right)
\sum_{p_1:~p_1\wedge k=1}
q^{\Delta}\bar{q}^{\bar{\Delta}
\left(\frac{p_1}{k},0,m,0\right)}.
\end{eqnarray}
as it should be.

From (\ref{zzz}), one can in principle read off the non-trivial selection rules in the bi-layer system.
In particular, it is always possible to calculate the lower spectrum by this explicit formula.
It seems, however, difficult for general cases to state the selection rules in simpler form e.g.
in terms of few quantum numbers. We do not pursue this direction, and just present some
examples in order to convey the flavor of how the selections occur.

\subsection{Spectrum in the bi-layered dilute/dense $O(1)$ models}\label{section_bilayero1}

The single-layer $O(1)$ model which is realized at $g=4\hat{g}=4/3$,
$e_0=1/3$
\cite{cgon} is the Ising model, which has the central charge $c=1/2$.
Hence, the bilayer model without interaction has $c=1$.
A partition function of CFT, in general, behave as 
\begin{eqnarray}
Z\sim\left(q\bar{q}\right)^{-\frac{c}{24}}.
\end{eqnarray}
By now, we use a combination which is suitable for analyzing $c=1$ partition functions:
\begin{eqnarray}
\mathcal{Z}_{c1}\left(\hat{g}, e_0; \Gamma\right)=
\left(q\bar{q}\right)^{-\frac{1}{24}}|\eta(q)|^4 Z,
\label{zbz}
\end{eqnarray}
which would start from the terms of order $1$ as the lowest-order term.
In general, if a conformal field theory has a primary fields of dimension $h$, the theory also has a descendant field of the dimension $h+n$  
which is, in the algebraic setting \cite{bpz}, obtained by applying 
the Virasoro generators
$L_{-k_i}$  $(k_i\geqslant 1)$ to the primary
such that $n=k_1+\cdots +k_m$. 
As is well known, this property is closely related to the fact that 
the inverse of the eta function (\ref{eta}) is almost 
the generating function of the partition numbers $d(n)$:
\begin{eqnarray}\label{etainverse}
\eta(q)^{-1}=q^{-\frac{1}{24}}\prod_{n=1}^\infty \left(1-q^n\right)^{-1}=
q^{-\frac{1}{24}}\sum_{n=0}^\infty d(n) q^n.
\end{eqnarray}
Correspondingly, we should note that the coefficients of 
$q^{h}\bar{q}^{\bar{h}}$
depends on the power of the eta function which we factored out.
However, here we focus on the lower spectrum in (\ref{zbz}) and 
do not concern the higher order terms corresponding to the descendants.
From now on, we study the coefficients which appear in (\ref{zbz}) and call them ``multiplicities".

Using (\ref{zzz}) with  (\ref{zoff}) and (\ref{zdiagoff}), we obtain
\begin{eqnarray}\label{decoupled}
\fl
\mathcal{Z}_{c1}\left(\frac{1}{3}, \frac{1}{3}; 0\right)=&\left(1-2 q+q^2\right)\left(1-2 \bar{q}+\bar{q}^2\right)
+q\bar{q}
+2\left( q^{\frac{1}{16}}- q^{\frac{17}{16}}\right)\left(\bar{q}^{\frac{1}{16}}-
\bar{q}^{\frac{17}{16}} \right) 
\nonumber\\&
+2 \left(q\bar{q}\right)^{\frac{9}{16}}
+\left(q\bar{q}\right)^{\frac{1}{8}}
+ 2 \left( q^{\frac{1}{2}}- q^{\frac{3}{2}}\right)\left(\bar{q}^{\frac{1}{2}}-\bar{q}^{\frac{3}{2}}\right)
+\mathcal{O}\left(q^2,\bar{q}^2\right),
\end{eqnarray}
where the symbol $\mathcal{O}\left(q^2,\bar{q}^2\right)$
represents the terms of higher-order than $q^2$ or $\bar{q}^2$.
The result (\ref{decoupled}) can be checked 
by using either the Ising partition function (\ref{ising}) or the orbifold one (\ref{orbifoldising}) in Appendix B.
Reflecting the non-interacting nature,
the spectrum is given by the trivial sum of two elements chosen from 
the single Ising spectrum $\{0,\,1/16,\,1/2\}$.

Next we turn to the interacting model.
A simple, and yet non-trivial spectrum is realized at $\lambda=-1$ ($\Gamma=3/2$). It reads, 
\begin{eqnarray}\label{gamma32}
\fl
\mathcal{Z}_{c1}\left(\frac{1}{3}, \frac{1}{3}; \frac{3}{2}\right)=&\left(1-q+q^2\right)\left(1-\bar{q}+\bar{q}^2\right)+
2\left(q\bar{q}\right)^{\frac{1}{16}}
+2 \left(q\bar{q}\right)^{\frac{9}{16}} 
+2 \left(q\bar{q}\right)^{\frac{17}{16}} 
\nonumber\\&
-2 \left(q\bar{q}\right)^{\frac{3}{8}} 
-2 \left(q\bar{q}\right)^{\frac{11}{8}}
+4\left( q^{\frac{1}{2}}- q^{\frac{3}{2}}\right)\left( \bar{q}^{\frac{1}{2}}- \bar{q}^{\frac{3}{2}}\right)
+\mathcal{O}\left(q^2,\bar{q}^2\right).
\end{eqnarray}
Comparing the interacting model (\ref{gamma32}) with 
the decoupled one (\ref{decoupled}),
we notice the presence of the new characteristic dimension $3/8$. 
This dimension also appears, for example, in the spectrum of
the level-$2$ coset model \cite{gko, date, richard} 
\begin{equation}
\frac{\widehat{s\ell}(2)_{2} \otimes \widehat{s\ell}(2)_2}{\widehat{s\ell}(2)_4},
\label{coset}
\end{equation}
or in the Ashkin-Teller quantum chain
at special points \cite{gehlen1,gehlen2,yang},
where the spectrum can be related to the irreducible representations
of the $c=1$, $\mathcal{N}=2$ superconformal algebra.  
The dimensions which appear in the table in the coset model are \cite{date, richard}
\begin{eqnarray}\label{cosetspectrum}
\left\{0,\,\frac{1}{24},\,\frac{1}{16},\,\frac{1}{6},\,
\frac{3}{8},\,\frac{9}{16},\,\frac{2}{3},\,1,\,\frac{3}{2}\right\}
\end{eqnarray}
Further, we remark that the three diagonal exponents
\footnote{In the $c=1$, $\mathcal{N}=2$
superconformal model,
the two exponents $1/24$ and $1/6$ belong
to the Ramond sector  
and 
to the Neveu-Schwarz sector, respectively \cite{qiu}.}
 contained in 
(\ref{cosetspectrum}) but not in (\ref{gamma32}), namely
$1/24$, $1/6$, and $2/3$
do appear also in the interacting model with $\hat{g}=1/6$,
$e_0=1/3$ $\left(\tilde{n}=1\right)$, and
$\Gamma=3$ $\left(\lambda=\exp \left[2\pi i\hat{g} \Gamma\right]=-1\right)$,
whose partition function reads,
\begin{eqnarray}\label{gamma3}
\fl
\eqalign{\mathcal{Z}_{c1}\left(\frac{1}{6}, \frac{1}{3}; 3\right)=
(q\bar{q})^{\frac{1}{24}}(3-4q-4\bar{q}+6q\bar{q})
-2\left[(q\bar{q})^{\frac{1}{6}}+(q\bar{q})^{\frac{2}{3}}\right]&(1-q-\bar{q})\\
&+\mathcal{O}\left(q^2,\bar{q}^2\right).}
\end{eqnarray}
The value $g=4\hat{g}=2/3$ also leads to $n=1$,
which is the same as in (\ref{gamma32}), but 
belongs to a different branch in (\ref{ng});
this value is known to correspond to {\it the dense} $O(1)$ model. 
The corresponding non-interacting model has the simpler function
\begin{eqnarray}\label{gamma0dense}
\fl
\mathcal{Z}_{c1}\left(\frac{1}{6}, \frac{1}{3}; 0\right)=
4\left( q \bar{q}\right)^{\frac{1}{24}}\left(1-2q-2\bar{q}+3q\bar{q}\right)
+\mathcal{O}\left(q^2,\bar{q}^2\right).
\end{eqnarray}
Here we just substitute the value for the dense model,  
although our argument (especially that in Section \ref{section_intersection}) is more reasonable
for the dilute model.
Note also that the lowest-order term in (\ref{gamma3}) which is diagonal in $q$ and $\bar{q}$ is
$\left( q \bar{q}\right)^{\frac{1}{24}}$, which means
in our notation (\ref{zbz}) that
(\ref{gamma3}) is related to a model with the central charge $c=0$
rather than to that with $c=1$ 
if it is considered as a full partition function. 
It remains, however, the possibility that  
(\ref{gamma3}) forms a part of some other partition function with the central charge  $c=1$.

In summary,  the partition functions of
the bi-layered dilute $O(1)$ models and 
the bi-layered dense $O(1)$ models 
both at the intersection weight $\lambda=-1$
are represented respectively by (\ref{gamma32}) and (\ref{gamma3}),
and together reproduce
all of the spectrum (\ref{cosetspectrum}) of the coset model (\ref{coset}).
It would be gratifying if one could find more precise relation
between the partition function of the coset model and 
special cases (which may include some of the examples in this section) of our modular invariant partition function. 

\subsection{Flow of the spectrum in the bi-layered dilute $O(1)$ model}\label{section_flow}

\begin{figure}[h!]
\begin{center}
\includegraphics[width=14cm]{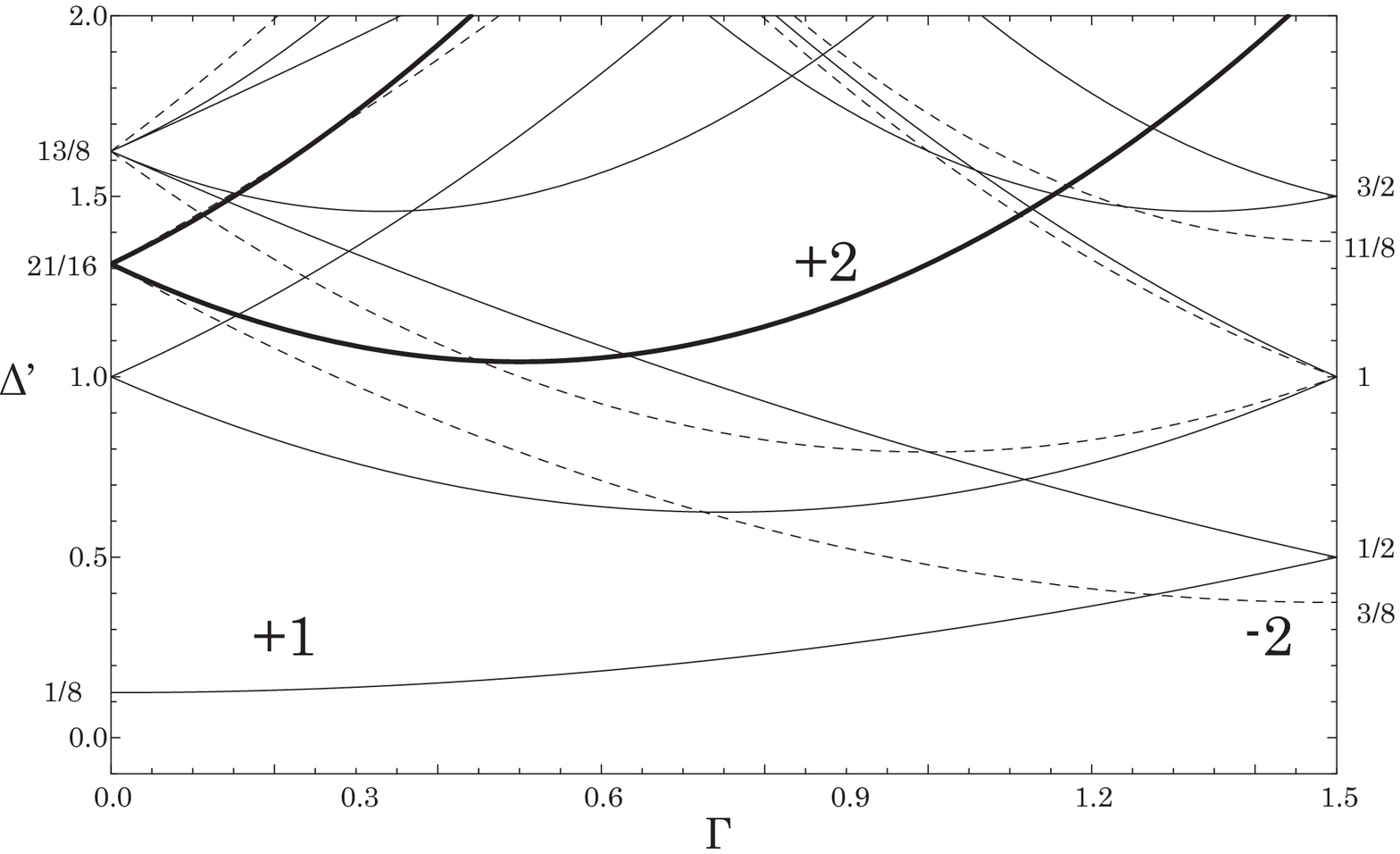}
\caption{Flow of the conformal dimensions $\Delta'=\Delta-1/24$ as functions of $\Gamma$ in 
the diagonal sectors of $\mathcal{Z}_{\neq 0}$ in the model with $g=1/3$
and $e_0=1/3$.
Each curve carries 
a definite multiplicity 
(solid: +1, thick: +2, dashed: -2).
}
\label{graph}
\includegraphics[width=14.2cm]{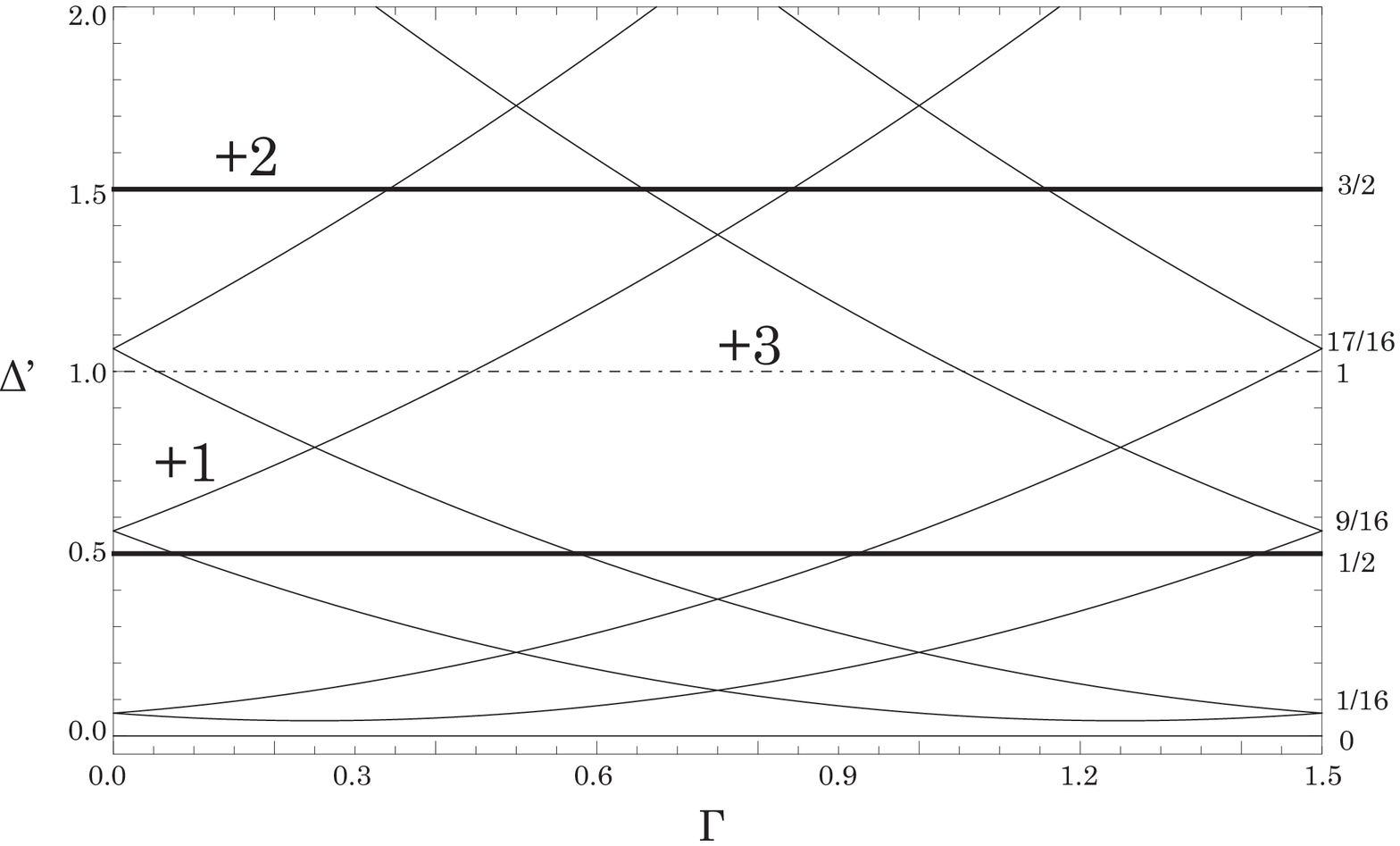}
\caption{Flow of the conformal dimensions as functions of $\Gamma$ in 
the diagonal sectors of $\mathcal{Z}_{0}$ in the same model ($g=1/3$
and $e_0=1/3$).
Again each curve carries 
a definite multiplicity 
(solid: +1, thick: +2, dot-dashed: +3).
}
\label{graph2}
\end{center}
\end{figure}

The curves of conformal dimensions belonging to the sector $\mathcal{Z}_{\neq 0}$ 
and $\mathcal{Z}_{0}$
connecting the dilute $O(1)$ model at $\lambda=1$ ($\Gamma=0$) and that at $\lambda=-1$ ($\Gamma=3/2$)
are plotted in Fig. \ref{graph} and Fig. \ref{graph2}, respectively.
As one can see from (\ref{cspin}), the spin $\Delta-\bar{\Delta}$ is independent of $\Gamma$.
Accordingly, we choose to show the dimensions of spin zero (diagonal) operators.
We plot $\Delta'=\Delta-1/24$ instead of $\Delta$ itself.
One can reproduce the diagonal terms 
in (\ref{decoupled}) and that in (\ref{gamma32}), including degeneracies,
by looking at the intersections of the curves to the left and to the right axis of both graphs.

All the curves in Fig. \ref{graph} are parabolas, as is clear from (\ref{cdimension}).
They are folded due to the symmetries of the partition function
that follow from the definition (\ref{lambda}):
\begin{equation}
\Gamma\to -\Gamma, \quad\quad
\Gamma\to \Gamma+3.
\label{gammasymmetry}
\end{equation}
In addition to these, as in Fig. \ref{graph2}, we have an extra symmetry 
\begin{equation}
\Gamma\to \Gamma+\frac{3}{2},
\label{gammahalfsymmetry}
\end{equation}
in the sectors $\mathcal{Z}_{0}$. 
Formally, this can be understood by the presence of 
the half-divided charge $e_0/2$ in (\ref{zdiagoff}).
The four flat lines ($\Delta'=0,1/2,1,$ and $3/2$)  correspond to
the operators with vanishing magnetic charges $m_A$ and $m_B$. 
As a result of (\ref{gammahalfsymmetry}), all the difference of the operator content of the model at $\lambda=1$
to that of the model at $\lambda=-1$ are contained in the sector $\mathcal{Z}_{\neq 0}$ (Fig. \ref{graph}).

Thus, let us concentrate on the flow in the sector $\mathcal{Z}_{\neq 0}$.
By recognizing that these parabolas occur in sequence,
we may state an interpretation of this flow as follows.
If we look at the graph up-side down,
these curves are regarded as some trajectories of balls
in a space surrounded by the walls at $\Gamma=0$ and at $\Gamma=3/2$
in the presence of a uniform gravitational field.
For example, if we are at $(\Gamma,\Delta')=(0,1/8)$ and
throw a ball in the horizontal direction to the wall at $\Gamma=3/2$, 
then it falls down to $\Delta'\rightarrow \infty$ 
as it reflects alternatively at $\Gamma=0$ and at $\Gamma=3/2$.   
If we tune the initial velocity, it goes as $(0,1/8)\to (3/2,1/2)\to (0,13/8)\to \cdots$, which in fact appears in the graph.
Another trajectory is $(3/2,3/8)\to (0, 21/16)\to \cdots$.
Every collisions are perfectly elastic  
due to the symmetry (\ref{gammasymmetry}). 
Hence, we have a law of the reflection in the spectrum.

Each falling trajectory carries a certain positive or negative multiplicity 
and implies the possibility of the operators of higher scaling dimensions at any $\Gamma$.
For fixed $\Gamma$, it defines a sequence of ``descendants''\footnote{Here, in general, difference of the dimensions
are restricted to neither integers nor half-integers;
they are not the usual descendants well-known in the conformal, or superconformal theories.}.  
At $\Gamma=0$, for instance,
the descendants of the field of dimension $\Delta'=1/8$ 
may have dimensions $\{13/8, 49/8, 109/8\cdots\}$. 
Finally, we remark that
cancellations of multiplicities may happen at particular values of $\Gamma$ where multiple trajectories cross each other.
As we can see from (\ref{decoupled}), there are no diagonal terms 
with $\Delta'=\bar{\Delta}'=13/8,$ or $21/16$. 
Complete cancellations in fact occur both
at $(0, 13/8)$ and at $(0,21/16)$.
This illustrates a non-trivial aspect of the selection rules.

\section{Conclusion and Discussions}\label{section_conclusion}
We have introduced a bi-layered loop model in which the inter-layer coupling depends on 
the signed number of intersections between loops living on the different layers. The model is first defined
microscopically on the lattice with the periodic boundary condition, and then the corresponding
continuum partition function is proposed based on the idea of the Coulomb gas. 
Within our definition of the inter-layer coupling, the local fluctuation decouples from the topological one. Accordingly, we are left with the sum over the topological sectors of two compact bosonic fields.  

The partition function has the modular invariance. After the Poisson resummation, we have the explicit expression which reveals the operator content. Each coefficient involves a combination of the Ramanujan sums, and contains non-trivial selection rules.   
We analyze few examples of the curves of scaling dimensions induced by the inter-layer coupling 
(Fig. \ref{graph} and \ref{graph2}). 
The layered model continuously changes itself 
from the decoupled model ($\lambda=1$) 
into the non-trivial coupled model ($\lambda=-1$), where the operator content is reorganized 
and the model has the simple spectrum (\ref{gamma32}).

In the definition of our layered model on the lattice, 
loops from different layers do not overlap but instead intersect each other. 
This feature is unique to our model. 
To compare, we mention the coupled loop model 
whose self-dual critical point was studied by algebraic methods
\cite{fendley}
and the models that arise from the replica approach to the disordered models
such as the random-bond $q$-state Potts model \cite{ludwig,dotsenko}
or the disordered $O(n)$ loop model \cite{shimada}.
In the former model, two Potts models, described respectively by 
two independent Temperley-Lieb algebras,
are considered on {\it the same} square lattice and coupled via the weight for the local overlap of loops. 
Similarly, for the latter models, the averaging procedures lead to the models of  
the replicated $M$-layers which are coupled through the overlap of loop segments (energy operators),  
that is, renormalized version of the bonds $J\vec{s}\cdot \vec{s}$ 
where $\vec{s}$ is the microscopic spin.
Although, at this stage, we could not relate our model and
these other coupled loop models, we briefly discuss a possible approach to 
the latter, disordered models by using the method in this paper.

In the replica method, the intermediate layered models were analyzed for general $M\in \mathbb{N}$ by conformal perturbation theories, and then $M\to 0$ limits should be taken. 
In this way, the two lines of the non-trivial fixed point emanating from 
the Ising point $q=2$ and $n=1$ are found
\cite{ludwig,dotsenko,shimada}. 
Along these critical lines, the effective central charge and, thermal and magnetic eigenvalues are estimated.
They are also numerically checked, for the random Potts case,
using the transfer matrix \cite{cardyjacobsen,chatelain}. 
Now our model can be generalized to the one consisting of $M$-layers.  
Since it would be again described by a quadratic action
similar to that used in this paper, it seems to be straightforward to obtain 
the partition function of the model for any $M\in \mathbb{N}$.
It would then give some insight on the replica limit if the analytic continuation in the variable 
$M$ and studying its $M\to 0$ limit could be possible at the
level of the exact partition function. 

Further important extension is within the bi-layered model
but is a generalization including several number of charges $e_i$ ($i=1,\cdots,N$), or equivalently, with several weights $\tilde{n}_i$ for non-contractible loops. 
Other than at $\tilde{n} \in \mathbb{N}$, this is necessary for obtaining integer coefficients.
In the single-layer case, 
$\{\tilde{n}_i\}$ is taken to be the set of 
the eigenvalues \cite{pasquier} of the adjacency matrices associated with the A-D-E Dynkin diagrams.
In this way,
it would be possible to examine whether we have a discrete series of the models similar to the unitary minimal model \cite{fqs} in the single-layer case or not.

Finally, we emphasize that the virtue of the CG method is its directness. 
Although we have only dealt with a specific case in multilayered system,
the continuum partition function was derived 
from the lattice formulation by a straightforward path-integral. 
We hope that our slight extension in two dimensions from single layer to multilayer will stimulate further 
progress in this direction.
 
\ack
I would like to thank S. Hikami for various suggestions and reading of the manuscript.
I am grateful to T. Kanazawa, T. Kimura, K. Sakai, T. Sasamoto, H. \& T. Shimada, H. Suwa, and T. Yoshimoto for helpful comments. 
I have benefitted from discussions with the 
participants in the 24th IUPAP international conference on statistical physics in July 2010.
I especially thank J. Cardy, J. L. Jacobsen, and R. M. Ziff for critical comments  and interests in the work.
I would also like to thank P. Zinn-Justin for sharing his insights on the dense loop models.
This research was supported by 
a Grant-in-Aid for the Japan Society for the Promotion of Science (JSPS) fellows. 
  
\appendix
\section{Derivation of the formula (\ref{appramanujan})}\label{section_derivation}

\begin{figure}[h!]
\begin{center}
\includegraphics[width=10cm]{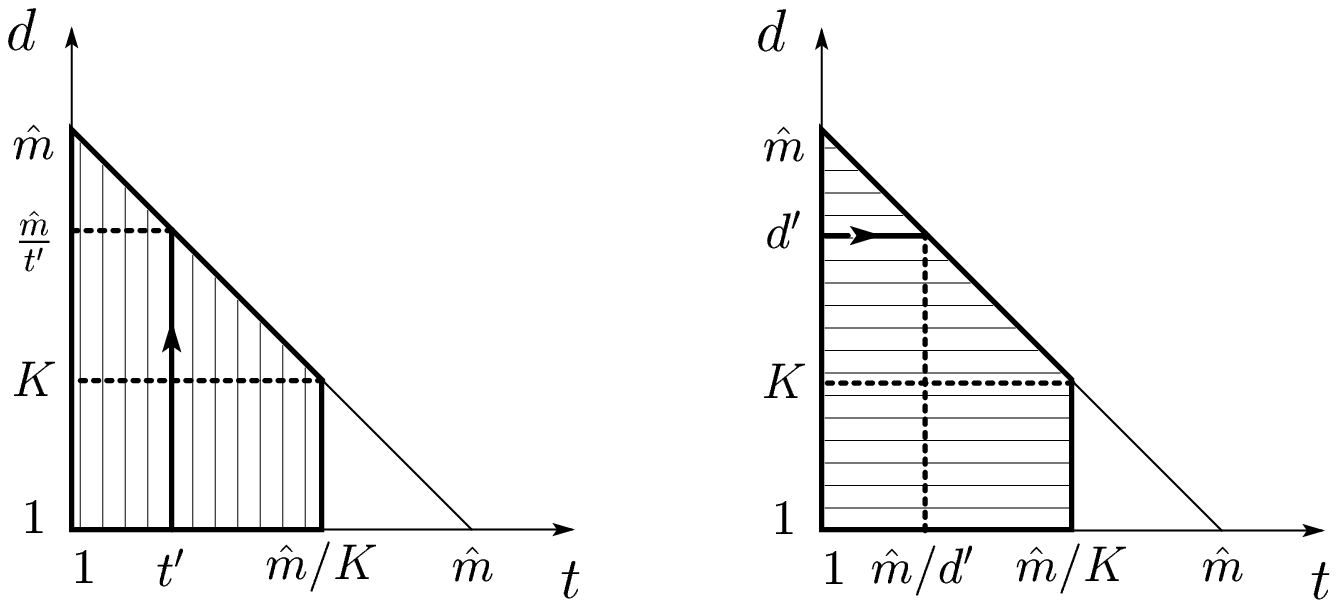}
\caption{The change of the order of the summations.
The summation is over the trapezoid in both figures;
a log-log scale is adopted so that the relation $td=\hat{m}$ holds
along the hypotenuse of the right triangle.}
\label{doublesum}
\end{center}
\end{figure}

We here present the derivation of the key formula (\ref{appramanujan}) for completeness although a similar calculation has been given in the literature \cite{readsaleur}.  
Let $F$ be the left hand side of (\ref{appramanujan}).
Changing the order of the summations and then rewriting by the reduced fraction $P/K$ yields
(below $\hat{m}$ denotes $|m|$ for readability),
\begin{eqnarray}
\fl
F&=&\sum_{k >0}\sum_{d |k }
\frac{\mu\left(\frac{k }{d }\right)}{k }
\cos \left(\pi e_0 d \right)
\sum_{\substack{m :~k |m\cr m\neq 0}}
\sum_{p \in \mathbb{Z}}
f\left(\frac{p}{k},m\right)\\
\fl
&=&
\sum_{m \neq 0}\sum_{k :~k |\hat{m} }
\sum_{d |k }
\frac{\mu\left(\frac{k }{d }\right)}{k }
\cos \left(\pi e_0 d \right)
\sum_{p \in \mathbb{Z}}
f\left(\frac{p}{k},m\right)\\
\fl
&=&
\sum_{m \neq 0}\sum_{K:~K|\hat{m} }\sum_{k :~K|k |\hat{m} }
\sum_{d |k }
\frac{\mu\left(\frac{k }{d }\right)}{k }
\cos \left(\pi e_0 d \right)
\sum_{P:~P\wedge K=1}
f\left(\frac{P}{K},m\right).
\end{eqnarray}
More convenient expression is obtained by setting $k=\hat{m}/t$: 
\begin{eqnarray}
\fl
F&=&\sum_{m \neq 0}\sum_{K:~K|\hat{m} }
\sum_{t:~K|\frac{\hat{m} }{t}}
\sum_{d:~d|\frac{\hat{m} }{t}}\frac{\mu\left(\frac{\hat{m} }{td}\right)}{\hat{m}/t}
\cos \left(\pi e_0 d \right)
\sum_{P:~P\wedge K=1}f\left(\frac{P}{K},m\right)\\
\fl
&=&\sum_{m \neq 0}\sum_{K:~K|\hat{m} }
\sum_{d:~d|\hat{m} }
\sum_{t:~t|\frac{\hat{m} }{K}\wedge \frac{\hat{m} }{d}}
\frac{\mu\left(\frac{\hat{m} /d}{t}\right)t}{\hat{m}}
\cos \left(\pi e_0 d \right)
\sum_{P:~P\wedge K=1}f\left(\frac{P}{K},m\right).
\label{afterchange}
\end{eqnarray}
In order to obtain (\ref{afterchange}), we have changed the order of the summation over $t$ and that over $d$;
the region of the summation is visualized in Fig. \ref{doublesum}.
After evaluating the sum over $t$ explicitly by applying the formula \cite{hardy} for the Ramanujan sum:
\begin{eqnarray}
\sum_{t:~t|(a\wedge b)}
\mu\left(\frac{a}{t}\right)t=\mathcal{C}_a(b),
\end{eqnarray}
we obtain the result in (\ref{appramanujan}). 

\section{The Ising and orbifold partition functions}\label{section_isingorbifold}

The partition function of the critical Ising model (the dilute $O(1)$ model) is, 
\begin{eqnarray}\label{ising}
Z_{\mathrm{Ising}}=\sum_{j=2}^4 \left|\frac{\theta_j}{2\eta}\right|
\end{eqnarray}
where $\theta_j$ $(j=2,3,4)$ are the classical theta functions
(at $z=0$) defined by
\begin{equation}\label{theta}
\fl\eqalign{
    \theta_{2}(z;\tau)
    & =
    \sum_{n \in \mathbb{Z}}
    q^{\frac{1}{2} \left( n + \frac{1}{2} \right)^2} \,
    \E^{2 \pi i \left( n+\frac{1}{2} \right) z}
	=\E^{\pi i z}q^{\frac{1}{8}}\prod_{n=1}^\infty\left(1-q^n\right)
\left(1+\E^{2\pi i z}q^n\right)\left(1+\E^{-2\pi i z}q^n\right)
,
    \\
    \theta_{3} (z;\tau)
    & =
    \sum_{n \in \mathbb{Z}}
    q^{\frac{1}{2} n^2} \,
    \E^{2 \pi \I  n  z}
	=\prod_{n=1}^\infty\left(1-q^n\right)
\left(1+\E^{2\pi i z}q^{n-\frac{1}{2}}\right)\left(1+\E^{-2\pi i z}q^{n-\frac{1}{2}}\right)
    ,
    \\
    \theta_{4} (z;\tau)
    & =
    \sum_{n \in \mathbb{Z}}
    q^{\frac{1}{2} n^2} \,
    \E^{2 \pi \I n \left( z+\frac{1}{2} \right) }
=\prod_{n=1}^\infty\left(1-q^n\right)
\left(1-\E^{2\pi i z}q^{n-\frac{1}{2}}\right)\left(1-\E^{-2\pi i z}q^{n-\frac{1}{2}}\right).}
\end{equation}
We need the square of (\ref{ising}). Here, it may be instructive to recall the partition function derived from 
the orbiforld boson (or that of the critical Ashkin-Teller model 
\cite{saleur_at, yang}), 
which includes the decoupled two-Ising models as a special point.
It is given by  
\begin{eqnarray}\label{orbifold}
Z_{\mathrm{orb}}(g)=\frac{1}{2}Z_{\mathrm{gauss}}(g,1)+\sum_{j=2}^4\left|\frac{\eta}{\theta_j}\right|
\end{eqnarray}
Formally, one can see the last sum corresponds to 
the cross term in the square of (\ref{ising}) by noting the identity between the eta function (\ref{eta}) and theta functions (\ref{theta}):
\begin{eqnarray}\label{etatheta234}
2\eta^3=\theta_2\theta_3\theta_4.
\end{eqnarray}
Physically, the sum in (\ref{orbifold}) has the meaning of the sum over the triplet of the spin structures (they exchange each other under the modular transformations) \cite{itzykson} 
\begin{eqnarray}
Z_{AP}=\left|\frac{\eta}{\theta_4}\right|,~
Z_{PA}=\left|\frac{\eta}{\theta_2}\right|,~
Z_{AA}=\left|\frac{\eta}{\theta_3}\right|,
\end{eqnarray}
where the subscripts corresponds to the boundary conditions along the two periods of the torus; $A$ and $P$ represent 
anti-periodic and periodic, respectively.
At the special point $g=1/2$, (\ref{orbifold}) indeed reduces to the square of (\ref{ising}):
\begin{eqnarray}\label{orbifoldising}
Z_{\mathrm{orb}}\left(g=\frac{1}{2}\right)=Z_{\mathrm{Ising}}^2,
\end{eqnarray}
as stated.

\end{document}